\documentclass[a4paper,fleqn]{cas-dc}
\usepackage[authoryear,longnamesfirst]{natbib}
\usepackage{graphicx}
\usepackage{algorithm}
\usepackage{algpseudocode}
\usepackage{soul}
\usepackage{amsmath,amsfonts}
\usepackage{color}

\def\tsc#1{\csdef{#1}{\textsc{\lowercase{#1}}\xspace}}
\tsc{WGM}
\tsc{QE}

\begin{document}
	\let\WriteBookmarks\relax
	\def\floatpagepagefraction{1}
	\def\textpagefraction{.001}
	\title [mode = title]{PSA Based Power Control for Cell-Free Massive MIMO under LoS/NLoS Channels}                      
	\author{\textcolor{black}{Ashish Pratap Singh}}
	\author{\textcolor{black}{Ribhu Chopra}}
	\affiliation{organization={Department of Electronics and Electrical Engineering, Indian Institute of Technology, Guwahati (emails: ashishpratapsingh196@outlook.com, ribhu@outlook.com)},
		postcode={781039}, 
		state={Assam},
		country={India}}

\begin{abstract}
 A primary design goal of the cell-free~(CF) massive MIMO architecture is to provide uniformly good coverage to all the user equipments~(UEs) connected to the network. However, it has been found that this requirement may not be satisfied in case the channels between the access points~(APs) and the UEs are mixed LoS/NLoS. In this paper, we try to address this issue via the use of appropriate power control in both the uplink and downlink of a CF massive MIMO system under mixed LoS/NLoS channels. We find that simplistic power control techniques, such as channel inversion-based power control perform sub-optimally as compared to max-min power control. As a consequence, we propose a particle swarm algorithm~(PSA) based power control algorithm to optimize the performance of the system under study. We then use numerical simulations to evaluate the performance of the proposed PSA-based solution and show that it results in a significant improvement in the fairness of the underlying system while incurring a lower computational complexity.  
\end{abstract}


\begin{keywords}
Cell-Free Massive MIMO \sep Line-of-Sight (LoS)  \sep Particle Swarm Algorithm (PSA) \
\end{keywords}

\maketitle

\section{Introduction}
Using a large number of antennas at the base station to serve a much smaller number of User Equipments (UEs) over the same time-frequency resources, known as massive multiple input multiple output~(mMIMO) has gained much research interest over the past decade and a half~[1]. It has been shown that subject to the availability of channel state information~(CSI) at the base station~(BS), mMIMO systems could offer high spectral  and energy efficiencies with simple signal processing~[2]. Due to these advantages, mMIMO has been widely accepted as a key enabling technology for the fifth generation of wireless systems~(5G)~[3, 4]. However, it has been shown that mMIMO fails to provide uniformly good coverage to all the UEs in the system~[5]. Consequently, a key goal in beyond 5G~(B5G) wireless systems is to provide uniformly good coverage to all the UEs, and it has been found that a distributed mMIMO system can achieve this goal~[6, 7]. One such architecture, dubbed cell free~(CF) mMIMO~[8] proposes a  distributed mMIMO system, wherein the BS is replaced with multiple access points~(APs) spread all across the area under service. These APs are connected to a central processing unit~(CPU) via a backbone network to serve all UEs over the same time-frequency resource~[9]. Since the UEs are now proximal to the APs, CF-mMIMO can provide a high likelihood of coverage~[10]. 

It is important to note that the high likelihood of coverage in a CF-mMIMO system implies a high likelihood of the existence of a line of sight~(LoS) path between the APs and the UEs. However, unlike the presently studied models in the literature that either consider an LoS link to be deterministically present~[11-13] or deterministically absent~[5, 9, 14-18], the presence of LoS link may depend on various physical parameters such as the physical locations of the APs and the UEs, the presence and density of the blockages, etc..~[19].  Therefore, the exact characterization and optimization of the performance of a CF-mMIMO system with probabilistic line-of-sight components become necessary~[20]. In this paper, our aim is to build on the previous work in this direction and develop power control techniques to optimize the performance of a CF-mMIMO system in the presence of probabilistic LoS components.  

The performance of CF-mMIMO systems under probabilistic LoS/NLoS channels has been discussed in detail for various receive architectures in~[20]. It has been shown that in the presence of probabilistic LoS channels, MMSE-based data detection achieves near-optimal performance, whereas conjugate beamforming-based data detection performs much worse even with the availability of accurate CSI at the CPU. It has also been shown that, at moderate AP densities, even under perfect/accurate CSI a fraction of UEs (approximately equal to the fraction of UEs without an LoS path) experience a SINR well below $0$ dB, indicating a ``capture" of the system by the UEs with LoS paths. This in turn implies the need for appropriate power control for all UEs to achieve equitable performance. 

In the context of power control in CF-mMIMO systems, the authors in~[9] have done an extensive comparative max-min fairness analysis of CF-mMIMO with centralized mMIMO. In~[5], a power control algorithm was developed to maximize energy efficiency considering hardware and backhaul power consumption. In~[21], authors exploited the max-min fairness optimization problem in rician fading by adapting the power and the AP–weighting receiver filter coefficients and  solves two sub-problems, receiver filter coefficients through a generalized eigenvalue problem and the power control problem by the bisection search method with linear programming. The authors in~[14] have considered max-min power control in rich scattering-based CF-mMIMO systems and developed a heuristic algorithm-based power control strategy that achieves near-optimal performance. In~[17], authors propose a bisection algorithm-based max-min power control strategy for downlink CF-mMIMO systems with rich scattering.

In most of the above-mentioned articles, the power allocation optimization problem is either solved through the sequential convex approximation (SCA) approach or by using convex approximation combined with the geometric programming (GP) approach. Since both approaches rely on second-order cone programming (SOCP), which requires high computational complexity and creates scalability issues for CF mMIMO systems, this way, to reduce computational complexity and run time, several works have investigated alternative optimization problems. The authors in~[22] solved the max-min optimization problem using a low-complexity method called as accelerated project gradient (APG). Furthermore, in~[23], a power allocation problem is solved for maximizing energy efficiency using the first-order method for non-convex programming. Additionally, the alternative optimization problem can also depend on deep learning. In~[24], authors performed the power optimization to maximize the network's sum rate using a heuristic sub-optimal scheme named as deep convolution neural networks (DCNN). Again, a deep learning-powered max-min-based uplink power allocation strategy was developed in~[25]. Further, in~[26], authors proposed a power control algorithm in the downlink, which solves the max-min UE fairness problem by modeling a deep neural network (DNN) with supervised training.  In~[27], it has been shown that using centralized and decentralized deep neural networks (DNNs) substantially reduces the complexity and processing time of the max-min power control optimization problem. Authors in~[28] solved the uplink max-min fairness power allocation optimization problem using meta-heuristics.

\begin{table}[t]
	\renewcommand{\arraystretch}{1.5}
	\caption[m1]{\normalsize{Notations used in paper}}
	\label{table:kysymys}\centering
	\begin{tabular}{|c|l|}
		\hline
		Notation & Definition \\ \hline
		$l_{m}$ & Height of the $m$-th AP   \\ 
		$l_{k}^{'}$  & Height of the $k$-th UE \\   
		$\delta_{mk}$ & Index variable for LoS channel between the \\
		& $m$-th AP and $k$-th UE \\
		$P_{mk}$ & Probability of existence of an LoS channel \\ & between 
		 the $m$-th AP and $k$-th UE \\
		$\bar{h}_{mk}$ & LoS channel gain between the $m$-th AP and \\ & $k$-th UE 	\\	
		$\dot{h}_{mk}$ & Fast fading coefficient for the NLoS channel \\ & between the $m$-th  AP and $k$-th UE \\
		$x_{mk}$ & $3$-D link distance between the $m$-th AP and \\ & $k$-th UE \\
		$d_{mk}$ & Horizontal ($2$-D) distance between the $m$-th \\ & AP and $k$-th UE \\
		$N$ & Number of antennas at the APs \\
		$d$ & Spacing between the antennas at all the APs \\
		$M$ & Number of APs in the system \\
		$K$ & Number of UEs in the system \\
		$d_{0}$ & Reference distance between any transceiver pair \\
		$\eta$ & Pathloss exponent \\
		$\alpha$ & Fraction of built-up area in the network \\
	    $\gamma$ & Average altitude of Building/blockages \\
		$\mu$ & Average number of blockages per unit area\\ 
		$\lambda_{c}$ & Carrier Wavelength \\
		$\mathbb{E}[.]$ & Expectation operation \\
		$\mathbf{0}_{N}$ & Null vector of length $N$ \\
		$\textbf{I}_{N}$ & Identity matrix of order $N$  \\ \hline
	\end{tabular}	
\end{table}
In this paper, we focus on improving the fairness of a CF-mMIMO system with mixed LoS/NLoS channels having moderate AP densities via power control. In this context, we first formulate the max-min power control problem for a CF-mMIMO system and demonstrate the optimality of its brute force solution in a two UE case. We then use one meta-heuristic, the particle swarm algorithm~(PSA), to obtain a lower complexity and with (near-)optimal solution for this problem. The primary motivation behind using the PSA algorithm is that it incurs lower computational complexity, low run time, and a (near) real-time adaptive approach that doesn't need offline training. These advantages of using PSA make it a viable, scalable, and efficient optimization technique to be employed in CF mMIMO networks. Our precise contributions are listed as follows.
\begin{enumerate}
	\item  We derive the rates achievable by the CF-mMIMO system with probabilistic LoS channels and use  Jain's Fairness Index to quantify the fairness for both uplink and downlink. We then use these to pose the power control problem for the underlying system.
	\item We consider a two UE test case and evaluate the performance of brute force-enabled max-min power control against channel inversion-based power control.
	\item Motivated by the significant gap in the performance of the two algorithms, and the high computational complexity of the brute force solution of the max-min power control, we develop a PSA-based power control strategy for the underlying system. 
	\item Via extensive numerical simulations, we show that our proposed PSA-based power control strategy results in near-optimal performance in the two UE case, and in significant improvement in the fairness in the 64 UE case as compared to channel inversion-based power control.  
\end{enumerate}
Therefore, the key takeaway of this work is that fairness can be achieved in CF-mMIMO systems with mixed LoS/NLoS using appropriate power control. Moreover, our proposed PSA approach can achieve near-optimal performance in this regard. 

The rest of the paper is organized as follows. Section~2 states the system and channel model. Section~3 describes our proposed system's performance analysis in both uplink and downlink. The description of Two UE Testing with performance comparison between channel inversion-based power control and without power control has been given in Section~4. Furthermore, in Section~5, the performance analysis of our proposed system using PSA-based power control has been done. In section~6, the computational complexity analysis of all considered power control strategies has been done. Next, the section~7 demonstrates all the simulation results with comprehensive discussions. In Section~8 brief conclusion to this paper has been made, and finally, in the end, references to the paper have been mentioned. The notation used in this paper is summarized in Table~\ref{table:kysymys},  and the simulation parameters used in this paper are listed in Table~ \ref{table:kysymysh}.

\section{System and Channel Models}
\begin{figure}[t]
	\centering
	\includegraphics[width=23em]{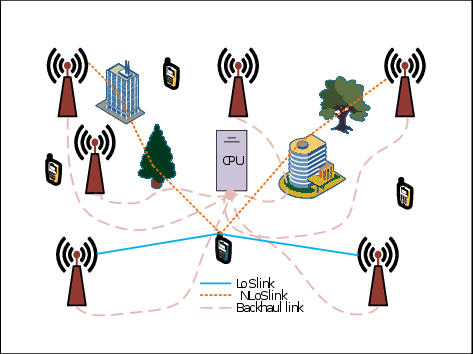}
	\caption{The System Model}
	\label{fig:e1}
\end{figure}
We consider a CF-mMIMO system comprising $M$ APs, each equipped with $N$ antennas, and the network serving a total of $K<<MN$ single antenna UEs. The heights of the $m$th AP and the $k$th UE are assumed to be $l_m$ and $l'_k$, respectively. This system operates in the time division duplexed~(TDD) mode, with an overall frame duration of $T=\tau_p+\tau_u+\tau_d$ channel uses or slots. Out of these, the first $\tau_p$ slots are utilized for CSI acquisition at the APs, the next $\tau_u$ slots are for uplink data transmission, and final $\tau_d$ slots are used for downlink data transmission. Similar to most other works on CF-mMIMO~[4, 8, 13], we assume the uplink and downlink channels to be reciprocal, negating the need for any downlink training. Also, for simplicity, we assume the pilot duration~($\tau_p$) to be long enough to avoid any pilot contamination and assume the availability of accurate CSI at all the APs. Here, we would like to note that the problem of CSI acquisition with limited pilots in a CF-mMIMO system with mixed LoS/NLoS channels is a non-trivial extension of the current literature on CSI acquisition~(see~[29] and the references therein) in CF-mMIMO systems and is beyond the scope of this work.
 \begin{table}[t]
	\renewcommand{\arraystretch}{1.5}
	\caption[m1]{\textsc{ Simulation Parameters}}
	\label{table:kysymysh}\centering
	\begin{tabular}{ |p{2.0cm}|p{3.1cm}|p{1.5cm}|  }
		\hline
		\textbf{Parameters}& \textbf{Descriptions} & \textbf{Values} \\
		\hline
		$d_{0}$ & Reference distance &$1$ m \\
		$f_{c}$ & Carrier Frequency   & $3.5$ GHz \\
		$l_{m}$ & AP antenna height & $10$ m \\
		$l'_{k}$    &UE antenna height & $1.5$ m \\
		$MN$ & AP antenna density & $1024/\text{km}^{2}$   \\
		$\mu$ & Average number of blockages per unit area  & $300/\text{km}^{2}$ \\
		$\alpha$ & Fraction of built-up area in the network & $0.5$ \\
		$\gamma$ & Average altitude of buildings/blockages & $20$ \\
		$\eta$ & Propagation Exponent & $3$ \\
        $P$ & Number of PSA Particles & $50$ \\
		$w$ & Weight Coefficient & $1$ \\
		$c_{1}$ & Cognitive Coefficient & $2$ \\
		$c_{2}$ & Social Coefficient & $2$ \\
		$Q$ & Number of iterations & $10000$ \\
		\hline
	\end{tabular}
\end{table}
We note that an LoS path may or may not exist between an AP and a UE. This is due to the random nature of the underlying blockages. Consequently, the channel between the $m$th AP and the $k$th UE, $\textbf{h}_{mk}\in\mathbb{C}^{N\times 1}$, is represented as,
\begin{equation}
\textbf{h}_{mk}=\delta_{mk}\bar{\textbf{h}}_{mk}+\sqrt{\beta_{mk}}\dot{\textbf{h}}_{mk},
\end{equation}
where  $\delta_{mk}$ is a binary-valued random variable that indicates the presence or the absence of an LoS component, $\bar{\textbf{h}}_{mk}$ denotes the LoS channel gain, $\beta_{mk}$ denotes the slow fading path loss component of the NLoS channel, and $\dot{\textbf{h}}_{mk}$ represents the fast fading component of the NLoS channel. More specifically, the fast fading NLoS channel, $\dot{\textbf{h}}_{mk}$, is assumed to consist of independent and identically distributed~(i.i.d) zero-mean circularly symmetric complex Gaussian (ZMCSCG) entries each having a unit variance, i.e., $\dot{\textbf{h}}_{mk}\sim\mathcal{CN}(0,\textbf{I}_{N})$. The slow fading coefficient of the NLoS component, $\beta_{mk}$, is modeled as $\beta_{mk}= \left(\frac{d_{mk}}{d_0}\right)^{-\eta}$, with $d_{mk}$ being the horizontal (two dimensional) distance between the $m$th AP and the $k$th UE and $d_0$ being the reference distance. On the other hand, $\delta_{mk}=1$ represents the presence of an LoS component and $\delta_{mk}=0$ indicates its absence, with Pr$\left\{\delta_{mk}=1\right\}=P_{mk}$. We model $P_{mk}$ according to the International Telecommunication Union~(ITU) blockage model~[29, 30] 
\begin{equation}
P_{mk}=(1-\omega)^{\sqrt{\alpha\mu d_{mk}}},
\end{equation}
where $\omega\triangleq\sqrt{\frac{\pi}{2}}\frac{\gamma}{l_{m}-l_{k}^{'}}\left[\text{erf}\left(\frac{l_{m}}{\gamma\sqrt{2}}\right)-\text{erf}\left(\frac{l_{k}^{'}}{\gamma\sqrt{2}}\right)\right]$, and erf(z)$\triangleq\frac{2}{\sqrt{\pi}}\int_{0}^{z}e^{-t^{2}}dt$,
$\gamma$ is the average altitude of blockages, $\alpha$ is the fraction of the built up area, and $\mu$ is the average number of blockages per unit area. Finally, the LoS channel gain between the $m$th AP and the $k$th UE takes the form
\begin{equation}
\bar{\textbf{h}}_{m,k}=\textbf{a}(\theta_{mk})\sqrt{G_{m}G_{k}}\left(\frac{l'_{k}l_{m}}{4\pi x_{mk}}\right)e^{\iota2\pi\frac{x_{mk}}{\lambda_{c}}},
\end{equation}
with $x_{mk}\triangleq\sqrt{d_{mk}^{2}+(l_{m}-l'_{k})^{2}}$ representing the three-dimensional distance between the respective APs and UEs, $\theta_{mk}$ being the azimuth angle between the AP and the UE, $\lambda_{c}$ denoting the carrier wavelength, $G_{m}$ and $G_{k}$ being the gains associated with the antennas at the $m$th AP and $k$th UE respectively, and $\textbf{a}(\theta$) representing the array response vector~(for an angle $\theta$) at the AP such that,
$	\textbf{a}(\theta)=[1,e^{\iota2\pi\frac{d}{\lambda_{c}}sin(\theta)},.........,e^{\iota2(N-1)\pi\frac{d}{\lambda_{c}}sin(\theta)}]^{T},
$
with $\iota=\sqrt{-1}$.

\section{Performance Analysis}

\subsection{Uplink}
We now evaluate the uplink performance of the system under consideration in the presence of accurate CSI at the CPU. We assume that all the UEs simultaneously transmit data to the APs over the same time-frequency resource and the $k$th UE transmits the symbol $q_k$, satisfying $\mathbb{E}\left[|q_k|^{2}\right]=1$. We can write the uplink signal received at the $m$th AP as,
\begin{equation}
\textbf{y}_{u,m} =\sum_{k=1}^{K}\sqrt{\zeta_{u,k}}\textbf{h}_{mk}q_{k}+\sqrt{N_{0}}\textbf{w}_{u,m},
\end{equation}
where $\zeta_{u,k}$ is the uplink power control coefficient for the $k$th UE, and $\textbf{w}_{u,m}\sim\mathcal{CN}(\textbf{0}_{N},\textbf{I}_{N})$. Letting  $\boldsymbol{\zeta_{u}}=[\zeta_{u,1},....,\zeta_{u,K}]^{T}$, $\textbf{q}=[q_{1},.....,q_{K}]^{T}$, $D_{\boldsymbol{\zeta}_u}=\text{diag}(\boldsymbol{\zeta}_u)$, $\textbf{H}_m=[{\mathbf{h}_{m1},\ldots,\mathbf{h}_{mK}}]$, and $\mathbf{H}=\left[\mathbf{H}^T_1,\ldots,\mathbf{H}^T_M\right]^T$, we can define the MMSE combining matrix at the CPU as, $\textbf{V}=(\textbf{HH}^{H}+N_{0}\textbf{I}_{MN})^{-1}\textbf{H}$. Then, letting $\textbf{y}_{u}=[\textbf{y}_{u,1},\textbf{y}_{u,2},.....,\textbf{y}_{u,M}]^{T}$, we can write the combined signal vector $\textbf{r} =\textbf{V}^{H}\textbf{y}_{u}$, at the CPU as,
\begin{multline}
\textbf{r}=\textbf{H}^{H}(\textbf{HH}^{H}+N_{0}\textbf{I}_{MN})^{-1}\textbf{H}\sqrt{D_{\boldsymbol{\zeta_{u}}}}\textbf{q}
\\+\sqrt{N_{0}}\textbf{H}^{H}(\textbf{HH}^{H}+N_{0}\textbf{I}_{MN})^{-1}\textbf{w},
\end{multline}
whose $k$th component, representing the $k$th UEs uplink data stream, can be written as,
\begin{align}
r_{k} &= \textbf{h}_{k}^{H}(\textbf{HH}^{H}+N_{0}\textbf{I}_{MN})^{-1}\textbf{h}_{k}\sqrt{\zeta_{u,k}}q_{k} \notag 
\\&\qquad+ \sum_{l=1,l\neq k}^{K}\textbf{h}_{k}^{H}(\textbf{HH}^{H}+N_{0}\textbf{I}_{MN})^{-1}\textbf{h}_{l}\sqrt{\zeta_{u,l}}q_{l} \notag 
\\&\qquad+ \sqrt{N_{0}}\textbf{h}_{k}^{H}(\textbf{HH}^{H}+N_{0}\textbf{I}_{MN})^{-1}\textbf{w} \\
&=f_{kk}\sqrt{\zeta_{u,k}}q_{k}+\sum_{l=1,l\neq k}^{K}f_{kl}\sqrt{\zeta_{u,l}}q_{l}+z_{u,k}.
\end{align}
Here, $f_{kk}=\textbf{h}_{k}^{H}(\textbf{HH}^{H}+N_{0}\textbf{I}_{MN})^{-1}\textbf{h}_{k}$ represents the effective channel coefficient for the $k$th UE's data stream, $f_{kl}=\textbf{h}_{k}^{H}(\textbf{HH}^{H}+N_{0}\textbf{I}_{MN})^{-1}\textbf{h}_{l}$ represents the interference channel coefficient from the $l$th UE's stream and $z_{u,k}=\sqrt{N_{0}}\textbf{h}_{k}^{H}(\textbf{HH}^{H}+N_{0}\textbf{I}_{MN})^{-1}\textbf{w}$ represents the  effective noise term.
Consequently, the uplink rate achievable by the $k$th UE can be written as,
\begin{equation}
R_{u,k}=\mathbb{E}\left[\log_{2}\left(1+\frac{|f_{kk}|^{2}
		\zeta_{u,k}}{\sum_{l=1,l\neq k}^{K}\mathbb{E}[|f_{kl}|^{2}]\zeta_{u,l}+\text{var}(z_{u,k})}\right)\right].
\end{equation}
Here, we need to choose the power control coefficients, $\zeta_{u,k}$, to optimize the system performance. Like most other works on cell-free massive MIMO~[16, 34, 35], which do not solve the expectations and use simulation techniques to evaluate the rates achievable by the system. We also do the same to evaluate the rates achievable by our considered system. To start with, we consider two approaches for power control, viz. channel inversion-based power control and max-min fairness-based power control, both are listed as follows, 
\subsubsection{Channel Inversion based Power Control}
\begin{figure}[t]
	\centering
	\includegraphics[width=27em]{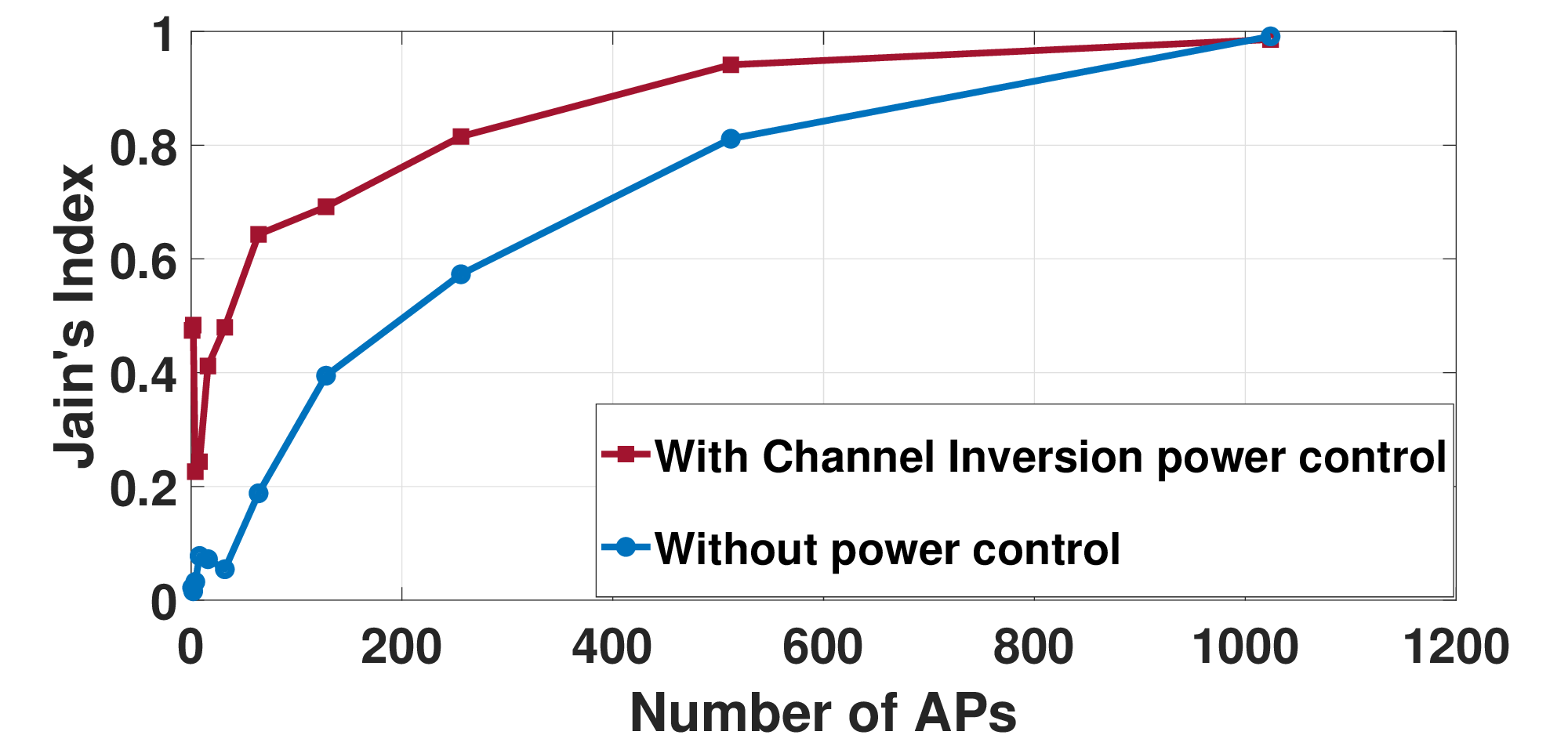}
	\caption{Comparison of Jain's Fairness Index  in uplink}
	\label{fig:e2}
\end{figure}

In this case, the uplink power control coefficient for the $k$th UE, $\zeta_{u,k}$, is defined as the inverse of the overall mean squared channel power, that is, 
\begin{equation}
\label{eqn:pcc_u}
\zeta_{u,k}=\frac{1}{\sum_{m=1}^{M}|\bar{\textbf{h}}_{mk}|^{2}+\sum_{m=1}^{M}\beta_{mk}} .
\end{equation} 
 This means that the UEs with better channel conditions would receive less power, while those with poor channel conditions would receive more power. This is a simple power control technique that has been used in cellular massive MIMO as an off the shelf solution, and acts as a good starting point for the problem of power control in cell free massive MIMO.
\subsubsection{Max-Min Power Control}
In this case, the uplink power control coefficients for all the UEs are chosen to maximize the minimum rate achieved by each UE, that is, $\zeta_{u,k}$ are chosen to satisfy
\begin{equation}
\begin{split}
&\underset{\zeta_{u,k}}{\text{max}}\hspace{0.5em}{\text{min}}\hspace{.5em}R_{u,k}   
\\
&\text{s.t} \hspace{1em}\zeta_{u,k}\geq 0 , \hspace{0.5cm}\forall\hspace{.5em} k=1,......,K.
\end{split}
\end{equation}

Finally, Jain's fairness index for the uplink rates can be expressed as 
\begin{equation}
J(R_{u})=\frac{\left(\sum_{k=1}^{K}R_{u,k}\right)^{2}}{K\sum_{k=1}^{K}(R_{u,k})^{2}}.
\end{equation}
 Here, the simulation setup considers $64$ UEs with both the UEs, and the APs uniformly distributed over a 1 $\text{km}^{2}$ area. We plot Jain's Fairness index for the system in the uplink as a function of the number of APs with and without channel inversion-based power control in Fig.~\ref{fig:e2}. Although, we can observe a dip in the system performance both with and without channel inversion based power control, for a smaller number of APs. This can be attributed to the availability of LoS paths to only a small number of users that ends up increasing the rate disparity among the users instead of alleviating it.
\subsection{Downlink}
We now evaluate the system performance in the downlink. Here, we assume that all $M$ APs serve all $K$ UEs simultaneously in the presence of accurate CSI at the CPU. Let $s_{k}$ be the intended downlink symbol for the $k$th UE, such that $\mathbb{E}\left[|s_k|^{2}\right]=1$, and $\textbf{s}=\left[s_{1},s_{2},....,s_{K}\right]^{T}$. Let $\textbf{A}_m=\left(\textbf{H}^{*}\textbf{H}^{T}+N_{0}\textbf{I}_{MN}\right)^{-1}\textbf{H}^{*}_m$ be the RZF precoding matrix at the $m$th AP, then the signal received by the $k$th UE in the downlink can be expressed as,
\begin{equation}
y_{d,k}=\sum_{m=1}^{M}\sqrt{\zeta_{d,k}}\textbf{h}_{mk}^{T}\textbf{A}_m\textbf{s}+\sqrt{N_{0}}w_{d,k}, \end{equation}
with  $\zeta_{d,k}$ being the downlink power control coefficient for the $k$th UE, and $w_{d,k}\sim\mathcal{CN}(0,1)$ being the AWGN. We can equivalently write,

\begin{align}
y_{d,k}&= \sqrt{\zeta_{d,k}}\textbf{h}_{k}^{T}\left(\textbf{H}^{*}\textbf{H}^{T}+N_{0}\textbf{I}_{MN}\right)^{-1}\textbf{h}_{k}^{*}s_{k} \notag\\
&\qquad+ \sum_{l=1,l\neq k}^{K}\sqrt{\zeta_{d,l}}\textbf{h}_{k}^{T}\left(\textbf{H}^{*}\textbf{H}^{T}+N_{0}\textbf{I}_{MN}\right)^{-1}\textbf{h}_{l}^{*}s_{l} \notag \\
&\qquad+ \sqrt{N_{0}}w_{d,k}. \\
&=\sqrt{\zeta_{d,k}}g_{kk}s_{k}+\sum_{l=1,l\neq k}^{K}\sqrt{\zeta_{d,l}}g_{kl}s_{l}+\sqrt{N_{0}}w_{d,k}.
\end{align}

Here, $g_{kk}=\textbf{h}_{k}^{T}\left(\textbf{H}^{*}\textbf{H}^{T}+N_{0}\textbf{I}_{MN}\right)^{-1}\textbf{h}_{k}^{*}$ represents the effective channel coefficient of the $k$th UE's desired signal, and $g_{kl}=\textbf{h}_{k}^{T}\left(\textbf{H}^{*}\textbf{H}^{T}+N_{0}\textbf{I}_{MN}\right)^{-1}\textbf{h}_{l}^{*}$ is the coefficient for the interference channel between the $k$th UE and the $l$th UE's data stream. Hence, the downlink rate achievable by the $k$th UE takes the form,
\begin{equation}
R_{d,k}=\mathbb{E}\left[\log_{2}\left(1+\frac{|g_{kk}|^{2}
		\zeta_{d,k}}{\sum_{l=1,l\neq k}^{K}\mathbb{E}[|g_{kl}|^{2}]\zeta_{d,l}+N_0}\right)\right].
\end{equation}
Similar to the uplink case, we can either use channel inversion-based power control or max-min fairness-based power control, with the values of $\zeta_{d,k}$ in the two cases being given as follows.
\subsubsection{Channel Inversion based Power Control}
\begin{figure}[t]
	\centering
	\includegraphics[width=27em]{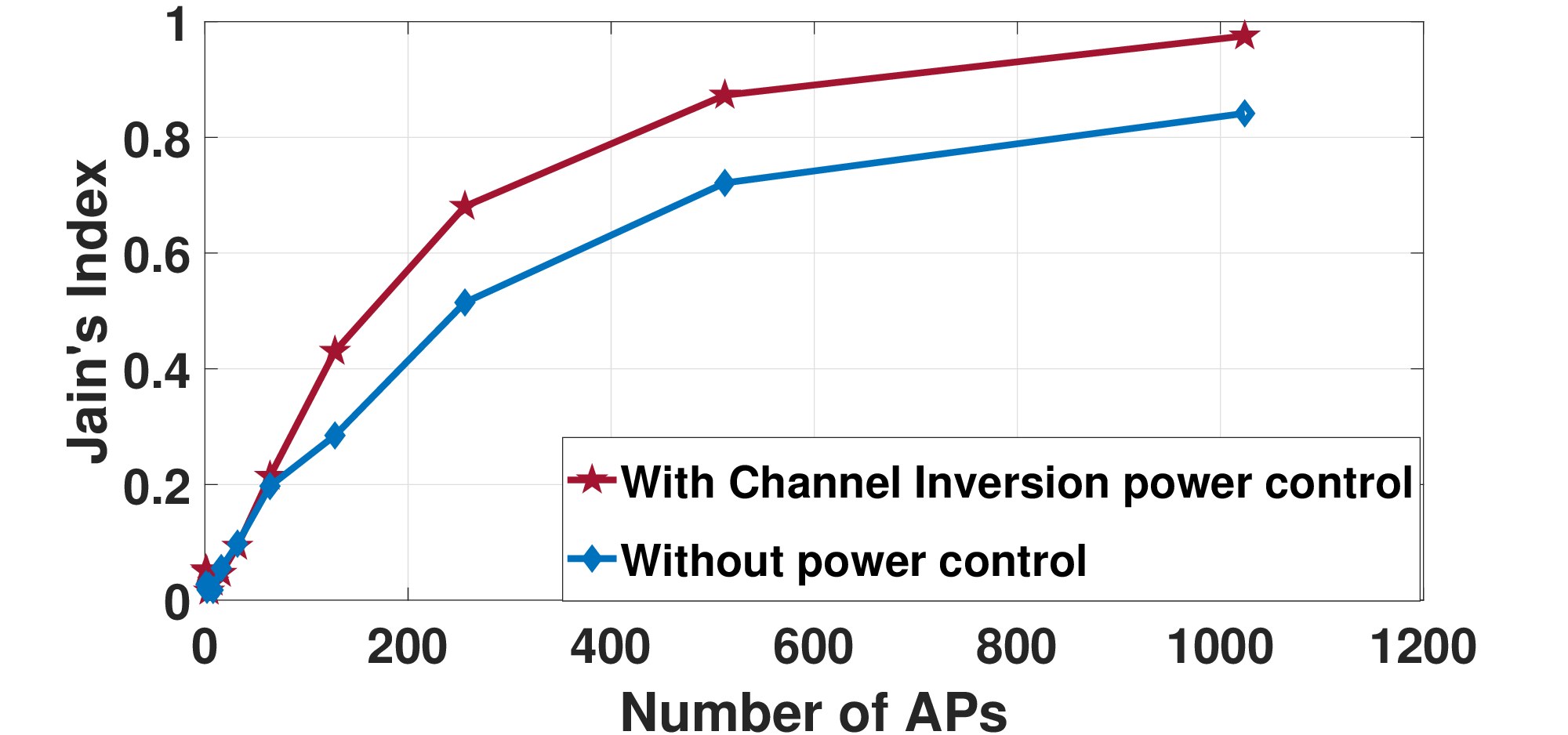}
	\caption{Comparison of Jain's Fairness Index  in downlink}
	\label{fig:e3}
\end{figure}

\begin{equation}
\label{eqn:pcc_d}
\zeta_{d,k}=\frac{1}{\sum_{m=1}^{M}|\bar{\textbf{h}}_{mk}|^{2}+\sum_{m=1}^{M}\beta_{mk}}.
\end{equation} 
\subsubsection{Max-Min Power Control}
$\zeta_{d,k}$ are chosen to satisfy
\begin{equation}
\begin{split}
&\underset{\zeta_{d,k}}{\text{max}}\hspace{0.5em}{\text{min}}\hspace{.5em}R_{d,k}   
\\
&\text{s.t} \hspace{1em}\zeta_{d,k}\geq 0 , \hspace{0.5cm}\forall\hspace{.5em} k=1,......,K.
\end{split}
\end{equation}
The consequent Jain's Fairness index in the downlink can be expressed as, 
\begin{equation}
J(R_{d})=\frac{\left(\sum_{k=1}^{K}R_{d,k}\right)^{2}}{K\sum_{k=1}^{K}(R_{d,k})^{2}}.
\end{equation}
We now plot Jain's fairness index both with and without power control in the downlink as a function of the number of APs  in Fig.~\ref{fig:e3}. The general trend of achieving a better fairness using channel inversion based power control motivates us to explore other power control algorithms. 
	\section{ Two UE Testing}
We now evaluate the performances of the two power control schemes under a test case with only two randomly placed UEs in the system.
\subsection{Uplink}
\begin{figure}[t]
	\centering
	\includegraphics[width=28em]{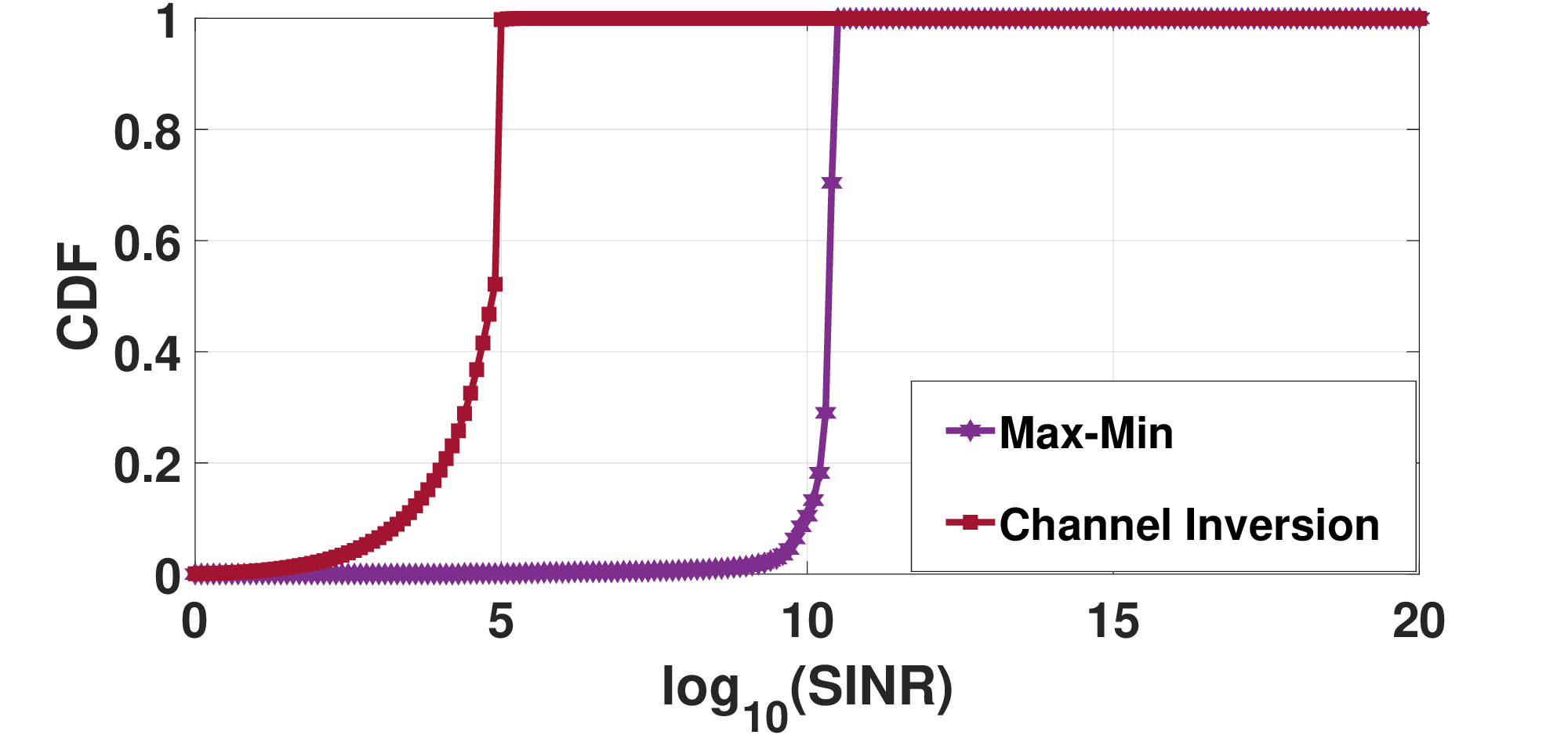}
	\caption{CDF of the minimum achievable uplink SINRs for different power control strategies.}
	\label{fig:e4}
\end{figure}

The uplink achievable rate for the first UE is,
The uplink achievable rate for the first UE is,
\begin{equation} 
R_{u,1}=\mathbb{E}\left[\log_{2}\left(1+\frac{|f_{11}|^{2}\zeta_{u,1}}{\mathbb{E}[|f_{12}|^{2}]\zeta_{u,2}+var(z_{u,1})}\right)\right], \\
\end{equation}
and for the second UE is,
\begin{equation}
R_{u,2}=\mathbb{E}\left[\log_{2}\left(1+\frac{|f_{22}|^{2}\zeta_{u,2}}{\mathbb{E}[|f_{21}|^{2}]\zeta_{u,1}+var(z_{u,2})}\right)\right],
\end{equation}
such that under channel inversion-based power control, 
\begin{equation}
\zeta_{u,1}=\frac{1}{\sum_{m=1}^{M}|\bar{\textbf{h}}_{m1}|^{2}+\sum_{m=1}^{M}\beta_{m1}} ,
\end{equation}
\begin{equation}
\zeta_{u,2} =\frac{1}{\sum_{m=1}^{M}|\bar{\textbf{h}}_{m2}|^{2}+\sum_{m=1}^{M}\beta_{m2}}.
\end{equation}

Similarly, under max-min power control $\zeta_{u,1}$ and $\zeta_{u,2}$ are chosen to satisfy
\begin{equation}
\begin{split}
&\underset{\left\{\zeta_{u,1},\zeta_{u,2}\right\}}{\text{max}}\hspace{0.1em}{\text{min}}\hspace{.5em}(R_{u,1},\hspace{.1em} R_{u,2}  )
\\
&\text{s.t} \hspace{1em}\zeta_{u,1}\geq 0 , \hspace{1em}\zeta_{u,2}\geq 0.
\end{split}
\end{equation}

 Considering a system having $256$ APs and $64$ single antenna UEs uniformly distributed over a 1 $\text{km}^{2}$ area operating at a carrier frequency of 2~GHz, we can now evaluate the relative performance of these two power control strategies. For this purpose, we have calculated the achievable SINRs for both cases over 10000 realizations. In Fig.~\ref{fig:e4}, we plot the CDF of minimum uplink SINR achievable by a UE. We observe that despite its simplicity, channel inversion-based power control is sub-optimal and results in inferior system performance, even with interference from just a single UE. This is mainly due to the fact that while channel inversion based power control, guarantees equal received SNR (on an average) to all the UEs it does not guarantee equal SINR, and hence fails. Max min power control, on the other hand has high complexity but guarantees a fixed minimum QoS to all the users.

\subsection{Downlink}
In this case, the rates achievable by the first and the second UEs are respectively given as,
\begin{equation}
R_{d,1}=\mathbb{E}\left[\log_{2}\left(1+\frac{|g_{11}|^{2}\zeta_{d,1}}{\mathbb{E}[|g_{12}|^{2}]\zeta_{d,2}+N_{0}}\right)\right],
\end{equation}
\begin{equation}
R_{d,2}=\mathbb{E}\left[\log_{2}\left(1+\frac{|g_{22}|^{2}\zeta_{d,2}}{\mathbb{E}[|g_{21}|^{2}]\zeta_{d,1}+N_{0}}\right)\right] .
\end{equation}
where
\begin{equation}
	\zeta_{d,1}=\frac{1}{\sum_{m=1}^{M}|\bar{\textbf{h}}_{m1}|^{2}+\sum_{m=1}^{M}\beta_{m1}}
\end{equation}
and  
\begin{equation}
\zeta_{d,2}=\frac{1}{\sum_{m=1}^{M}|\bar{\textbf{h}}_{m2}|^{2}+\sum_{m=1}^{M}\beta_{m2}} ,
\end{equation}
for channel inversion-based power control.
Similarly, under max-min power control $\zeta_{d,1}$ and $\zeta_{d,2}$ are chosen to satisfy 
\begin{equation}
\begin{split}
&\underset{\left\{\zeta_{d,1},\zeta_{d,2}\right\}}{\text{max}}\hspace{0.1em}{\text{min}}\hspace{.5em}(R_{d,1},\hspace{.1em} R_{d,2}  )
\\
&\text{s.t} \hspace{1em}\zeta_{d,1}\geq 0 , \hspace{1em}\zeta_{d,2}\geq 0.
\end{split}
\end{equation}

\begin{figure}[t]
	\centering
	\includegraphics[width=27em]{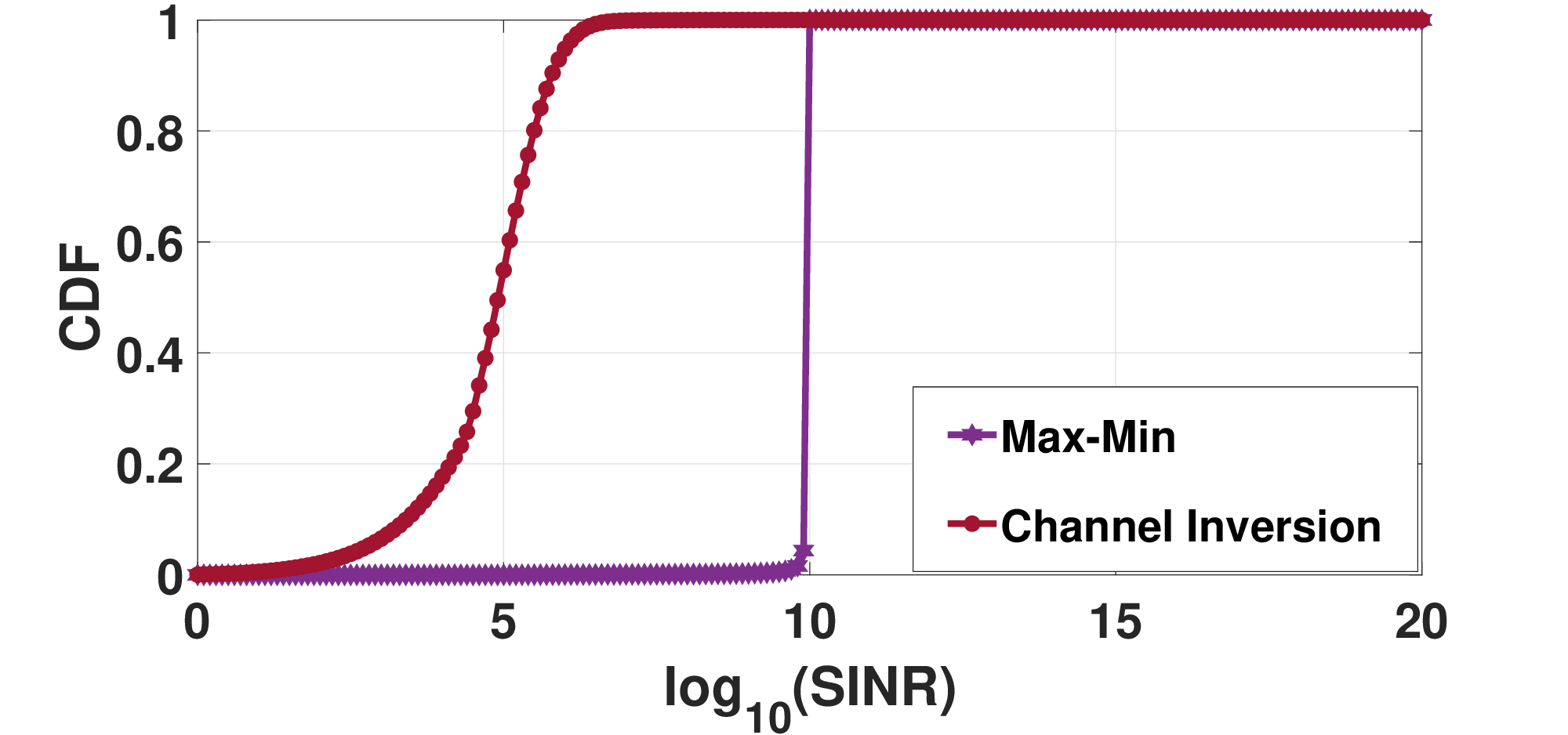}
	\caption{CDF of the minimum achievable downlink SINRs for different power control strategies.}
	\label{fig:e5}
\end{figure}

We also evaluate the relative performance of the two power control strategies in downlink. We plot the comparison of the CDFs of the minimum achievable downlink SINRs for a UE over 10000 realizations in Fig.~\ref{fig:e5}. We observe that channel inversion-based power control is sub-optimal in downlink as well. Again, a gap between the SINRs achievable in both uplink and downlink under the two power control strategies can be observed, necessitating the development of low-complexity power control techniques for the underlying system model. In the next section, we describe the use of the particle swarm algorithm as a candidate solution to our problem.   

\section{Particle Swarm Algorithm Based Power Control}
\begin{algorithm}[t]
	\caption{PSA based power control for a CF-mMIMO system under LoS/NLoS channel}\label{Alg_1}
	\begin{algorithmic}
		\State Initialize $\mathbf{X}^{gbest}=0$, val($\mathbf{X}^{gbest}$) = 0
		\For  {$p$=$1$ to $P$}
		\State Initialize  $\mathbf{X}^{p}\sim(\mathcal{U}[0,1])^K$, $\mathbf{V}^{p}\sim(\mathcal{U}[0,1])^K$
		\State $\text{val}(\mathbf{X}^{p})=g(\textbf{X}^{p})$
		\State $\mathbf{X}^{pbest(p)}=\mathbf{X}^{p}, \hspace{0.5em} \text{val}(\mathbf{X}^{pbes(p)})=\text{val}(\mathbf{X}^{p})$
		\If{$\text{val}(\mathbf{X}^{pbest(p)})>\text{val}(\mathbf{X}^{gbest})$}
		\State  $\text{val}(\mathbf{X}^{gbest})=\text{val}(\mathbf{X}^{pbest(p)})$
		\State $\mathbf{X}^{gbest}=\mathbf{X}^{pbest(p)}$
		\EndIf
		\EndFor
		
		\For{t = $1$ to $Q$}
		\For{$p$=$1$ to $P$}
		\State $\mathbf{V}^{p}(t+1)=w\mathbf{V}^{p}(t)+c_{1}\mathbf{r}_{1}\odot\left(\mathbf{X}^{pbest(p)}(t)-\mathbf{X}^{p}(t)\right)+c_{2}\mathbf{r}_{2}\odot(\left(\mathbf{X}^{gbest}(t)-\mathbf{X}^{p}(t)\right)$
		\State $\mathbf{X}^{p}(t+1)=\mathbf{X}^{p}(t)+\mathbf{V}^{p}(t+1)$
		\State $ \text{val}(\mathbf{X}^{p}(t+1))=g(\mathbf{X}^{p}(t+1))$
		\If {$\text{val}(\mathbf{X}^{p}(t+1))>\text{val}(\mathbf{X}^{pbest(p)})(t)$}
		\State $\mathbf{X}^{pbest(p)}(t)=\mathbf{X}^{p}(t+1)$
		\State $\text{val}(\mathbf{X}^{pbest(p)}(t))=\text{val}(\mathbf{X}^{p}(t+1))$
		\EndIf
		\If{$\text{val}(\mathbf{X}^{pbest(p)}(t))>\text{val}(\mathbf{X}^{gbest}(t))$}
		\State  $\text{val}(\mathbf{X}^{gbest})=\text{val}(\mathbf{X}^{pbest(p)}(t))$
		\State $\mathbf{X}^{gbest}(t)=\mathbf{X}^{pbest(p)}(t)$
		\EndIf
		\EndFor
		\EndFor
	\end{algorithmic}
\end{algorithm}

Particle Swarm Algorithm~(PSA) is a bio-inspired algorithm that solves an optimization problem by first generating a population of candidate solutions (particles) and then iteratively improving each particle's performance~[32, 33]. Here we consider a total of `$P$' particles, such that the $p$th particle's location is given by the vector $\mathbf{X}^{p}$ and velocity is given by $\mathbf{V}^{p}$. The algorithm's operation can be summarized as follows, the $p$th particle travels through the solution space of power control coefficients based on the sum of two velocity vectors. These velocity vectors are in the direction of the local best solution of that particle and the global best solution seen by the algorithm in general. 

To elaborate, the objective function to be optimized, known as the fitness function, is evaluated for each particle at each instant of time. Following this, each particle stores the coefficients achieving the best value of the fitness function as $\mathbf{X}^{pbest(p)}$ and the corresponding value is given as $\text{val}(\mathbf{X}^{pbest(p)})$. Similarly, the position corresponding to the best value by all the particles over all time is stored as $\mathbf{X}^{gbest}$, and the corresponding value of the fitness function is stored as~$\text{val}(\mathbf{X}^{gbest})$. 

Following this, in each iteration, and for each particle we generate two velocity vectors, one pointing towards the vector $\mathbf{X}^{pbest(p)}(t)$, and another towards $\mathbf{X}^{gbest}(t)$. The velocity of each particle is then updated using a linear combination of its current velocity with these two vectors, that is,

\begin{multline}
	\mathbf{V}^{p}(t+1)=w\mathbf{V}^{p}(t)+c_{1}\mathbf{r}_{1}\odot\left(\mathbf{X}^{pbest(p)}(t)-\mathbf{X}^{p}(t)\right)
	\\+c_{2}\mathbf{r}_{2}\odot(\left(\mathbf{X}^{gbest}(t)-\mathbf{X}^{p}(t)\right) 
\end{multline}
with $\mathbf{r}_{1}\sim\mathcal{U}[0,1]$ and $\mathbf{r}_{2}\sim\mathcal{U}[0,1]$. 

 Additionally, the movement in each direction is weighted by a different random factor to provide a higher degree of freedom. Following this, the position of each particle is updated as,
\begin{equation}
\mathbf{X}^{p}(t+1)=\mathbf{X}^{p}(t)+\mathbf{V}^{p}(t+1).
\end{equation}
 In Algorithm~\ref{Alg_1}, $\odot$ represents the element-wise product of two vectors. We use $w$ to denote the inertia weight coefficient, $c_{1}$ and $c_{2}$ are the  cognitive and social coefficients, respectively, which act as weights of the stochastic term regarding the difference vectors between the particle position. 
\subsection{Uplink Power Control}
We can write the minimum rate achievable by a UE as 
\begin{equation}
\Omega(\boldsymbol{\zeta}_{u})={\text{min}}\hspace{0.5em}\Psi_{k}(\boldsymbol{\zeta}_{u}),
\end{equation}
where $\boldsymbol{\zeta}_{u}=[\zeta_{u1},....,\zeta_{u,K}]^{T}\in \left(\mathcal{R}^{K}\right)^{+}$ and 
\begin{equation*}
\Psi_{k}(\boldsymbol{\zeta}_{u})=\left[\log_{2}\left(1+\frac{|f_{kk}|^{2}\zeta_{u,k}}{\sum_{l=1,l\neq k}^{K}\mathbb{E}[|f_{kl}|^{2}]\zeta_{u,l}+\text{var}(z_{u,k})}\right)\right].
\end{equation*}
Now, the optimization problem for providing equitable uplink rates to all the UEs in the system can be stated as,
\begin{align}
\label{eqn:opti_problem_u}
\begin{split}
&\underset{\boldsymbol{\zeta}_{u}}{\text{max}}\hspace{0.5em}\Omega(\boldsymbol{\zeta}_{u})
\\
&\text{s.t} \hspace{1em}\boldsymbol{\zeta}_{u} \geq 0.
\end{split}
\end{align}

We can solve~\eqref{eqn:opti_problem_u}, using PSA, according to Algorithm~\ref{Alg_1}.
\subsection{Downlink Power Control}
The minimum downlink rate achievable can be written as
\begin{equation}
\Omega(\boldsymbol{\zeta}_{d})={\text{min}}\hspace{0.5em}\Psi_{k}(\boldsymbol{\zeta}_{d}),
\end{equation}
where $\boldsymbol{\zeta_{d}}=[\zeta_{d1},....,\zeta_{d,K}]^{T}\in \left(\mathcal{R}^{K}\right)^{+}$ and $f_{k}(\boldsymbol{\zeta}_{d})$ is given by,
\begin{equation*}
\Psi_{k}(\boldsymbol{\zeta}_{d})=\left[\log_{2}\left(1+\frac{|g_{kk}|^{2}\zeta_{d,k}}{\sum_{l=1,l\neq k}^{K}\mathbb{E}[|g_{kl}|^{2}]\zeta_{d,l}+N_{0}}\right)\right].
\end{equation*}
The optimization problem for providing equitable downlink rates to all the UEs in the system can be stated as \\
\begin{align}
\label{eqn:opti_problem_d}
\begin{split}
&{\underset{\boldsymbol{\zeta}_{d}}{\text{max}}\hspace{0.5em}\Omega(\boldsymbol{{\zeta}}_{d})}
\\
&\text{s.t} \hspace{1em}\boldsymbol{\zeta}_{d} \geq 0 ,
\end{split}
\end{align}
We can solve~\eqref{eqn:opti_problem_d}, using PSA, according to algorithm~\ref{Alg_1}, similar to the uplink case.

\section{Computational Complexity}
In this section, we provide the computational complexity analysis of the three power control techniques discussed previously, viz. channel inversion based power control, max-min fairness based power control, and our proposed PSA based power control.
 \begin{table}[t]
	\renewcommand{\arraystretch}{1.6}
	\caption[m1]{\textsc{Computational Complexity of different Power Control Strategies}}
	\label{table:cca}\centering
	\begin{tabular}{ |p{3.5cm}|p{3.7cm}|  }
		\hline
		\textbf{Power Control Strategy}& \textbf{Computational Complexity} \\
		\hline
		Channel Inversion &  $\mathcal{O}(MK)$ \\
		Max-Min &  $\mathcal{O}(L^{K})$  \\
		PSA & $\mathcal{O}(M^{3}KP)$   \\
		\hline
	\end{tabular}
\end{table}
\subsection{Channel Inversion based Power Control}
In this case, the calculation of power control coefficients using equations~\eqref{eqn:pcc_u} and equation~\eqref{eqn:pcc_d} requires approximately $MK$ operations, resulting in an overall computational complexity given by $\mathcal{O}(MK)$. 
\subsection{Max-Min Power control}
 We have considered here a brute-force enabled max-min fairness based power control, which involves an exhaustive search over all possible power allocations to find an optimal solution. The key step for the power control algorithm is the enumeration of all the possible power levels. In our approach, we search over $L$ possible power levels~(within the power constraint) per UE, for the $K$ UEs, resulting in a search over $L^{K}$ possible power levels. This results in the computational complexity for max-min power control being $\mathcal{O}(L^K)$.
 \subsection{PSA based Power Control}  
 The computational complexity of Particle Swarm Algorithm~(PSA) based power control primarily depends on the number of particles, the dimensionality of search space, evaluation of fitness function and the number of iterations~[28]. The evaluation of fitness functions~\eqref{eqn:opti_problem_u} and~\eqref{eqn:opti_problem_d} for $P$ particles requires approximately $M^{3}KP$ operations. Following this, the update of particle positions and velocities requires $KP$ operations. Consequently, the computational complexity per iteration is $\mathcal{O}\left(M^{3}KP\right)$. 
 
\section{Simulation Results}
In this section, we present simulation results to evaluate the performance of our proposed PSA-based power control approach. For the purpose of these simulations, we consider $M$ APs and $K$ single-antenna UEs distributed uniformly over a 1 $\text{km}^{2}$ area. We assume that the NLoS propagation constant $\beta_{mk}$ to take the form $\beta_{mk} = \text{min}(1,( x_{mk}/d_{0} )^{-\eta})$, with $d_{0}$  being the reference distance, and $\eta$ being the propagation exponent. Unless stated otherwise, Table \ref{table:kysymysh} lists all the parameters used for these experiments.

\subsection{The Two UE Case}
\begin{figure}[t]
	\centering
	\includegraphics[width=27em]{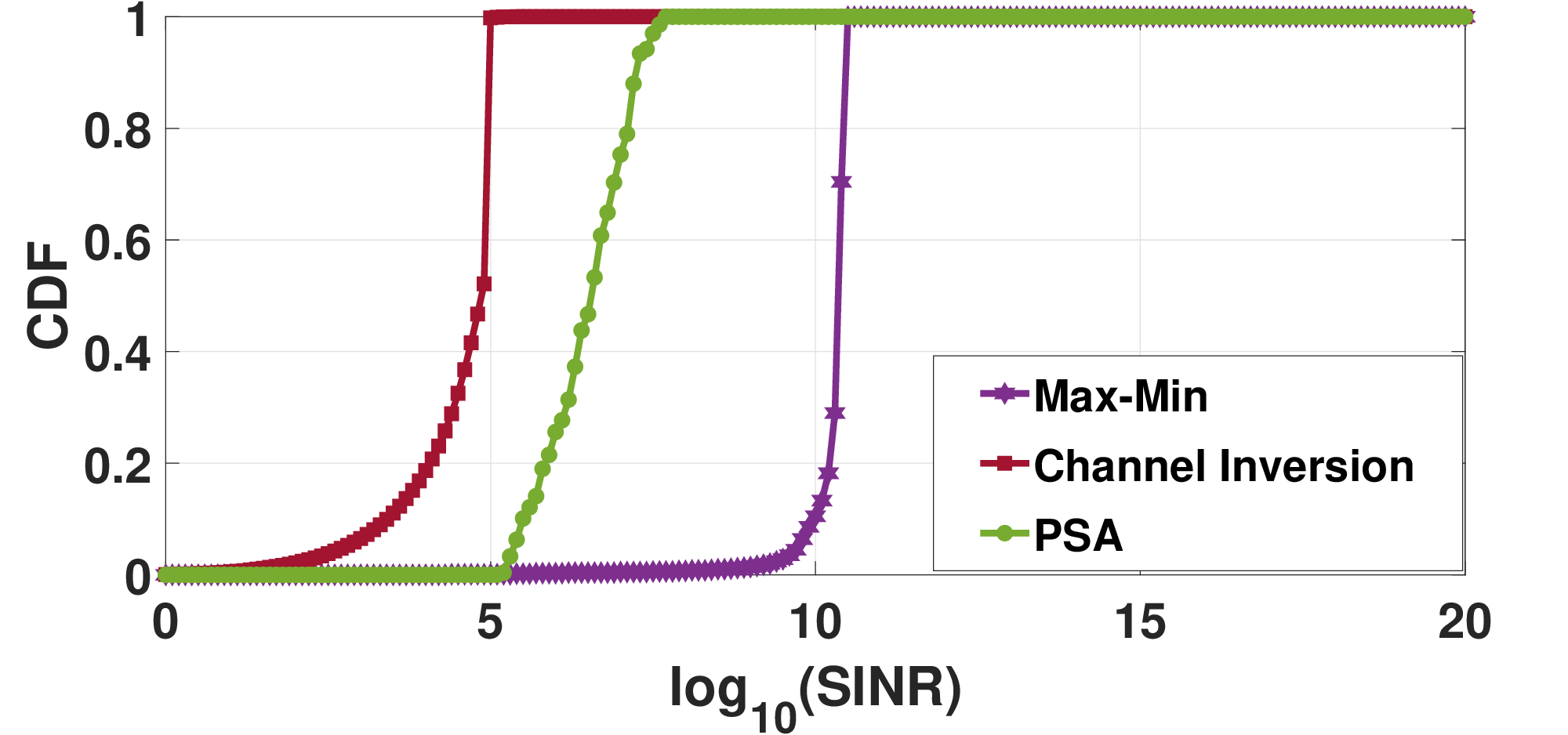}
	\caption{CDF of the minimum achievable uplink SINRs for different power control strategies in $2$ UEs case.}
	\label{fig:e6}
\end{figure}

\begin{figure}[t]
	\centering
	\includegraphics[width=29em]{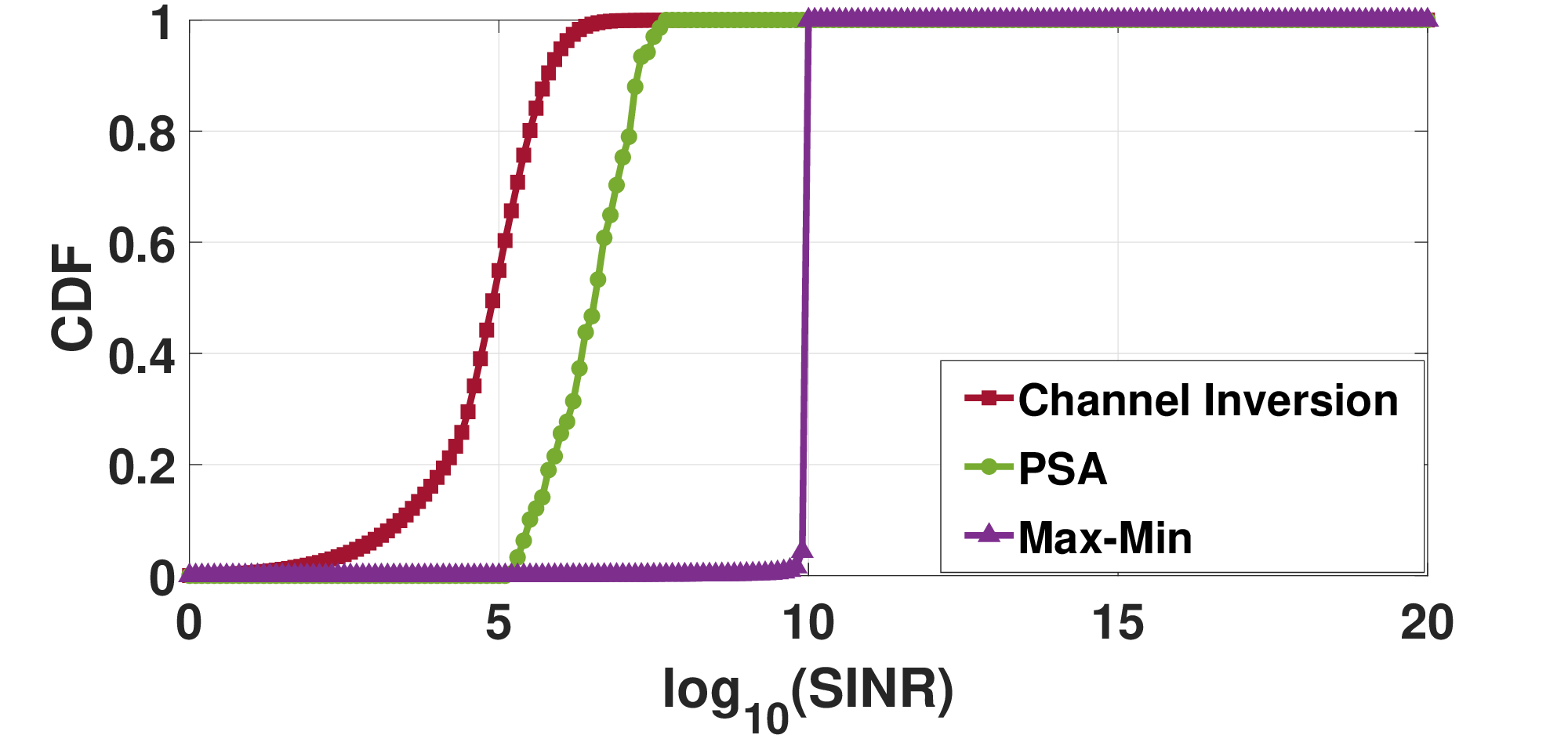}
	\caption{CDF of the minimum achievable downlink SINRs for different power control strategies in $2$ UEs case.}
	\label{fig:e7}
\end{figure}

We first compare the CDFs of the minimum achievable rates of the system having $256$ APs and $2$ UEs under the three power control strategies discussed in this paper for both uplink and downlink in Fig.~\ref{fig:e6} and Fig.~\ref{fig:e7} respectively. After keen observation, we see that by the use of PSA-based power control, the 90\% likely minimum achievable rate of the considered system can be improved by about 60\% in both uplink downlink directions as compared to channel inversion-based power control. This results in a performance closer to the max min-based power control. This significant performance gap between channel inversion and PSA-based power control motivates the use of PSA-based power control for a higher number of UEs connected in the network. 


\subsection{The 64 UE case}
\begin{figure}[t]
	\centering
	\includegraphics[width=27em]{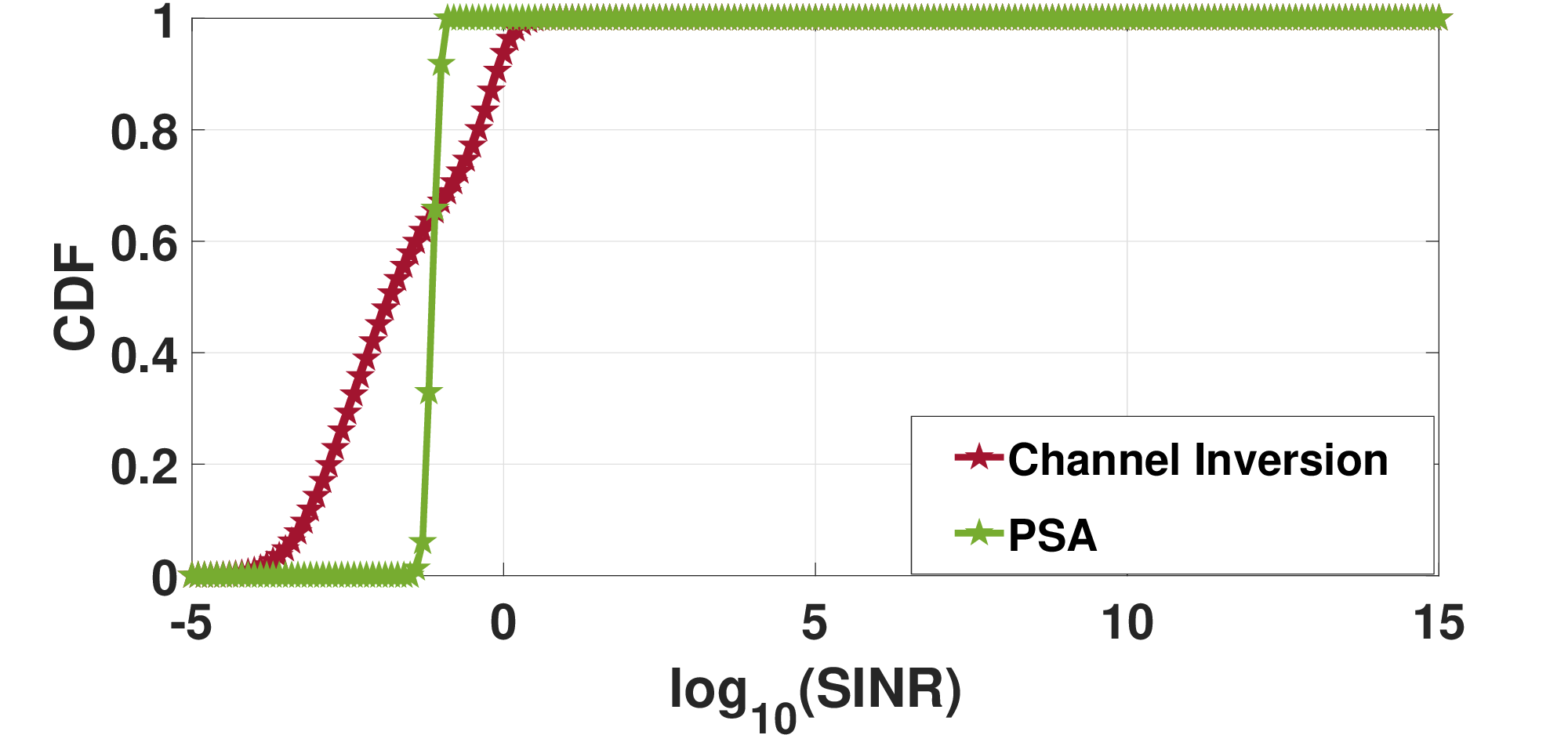}
	\caption{CDF of the minimum achievable uplink SINRs for different power control strategies in $64$ UEs case.}
	\label{fig:e8}
\end{figure}

\begin{figure}[t]
	\centering
	\includegraphics[width=27em]{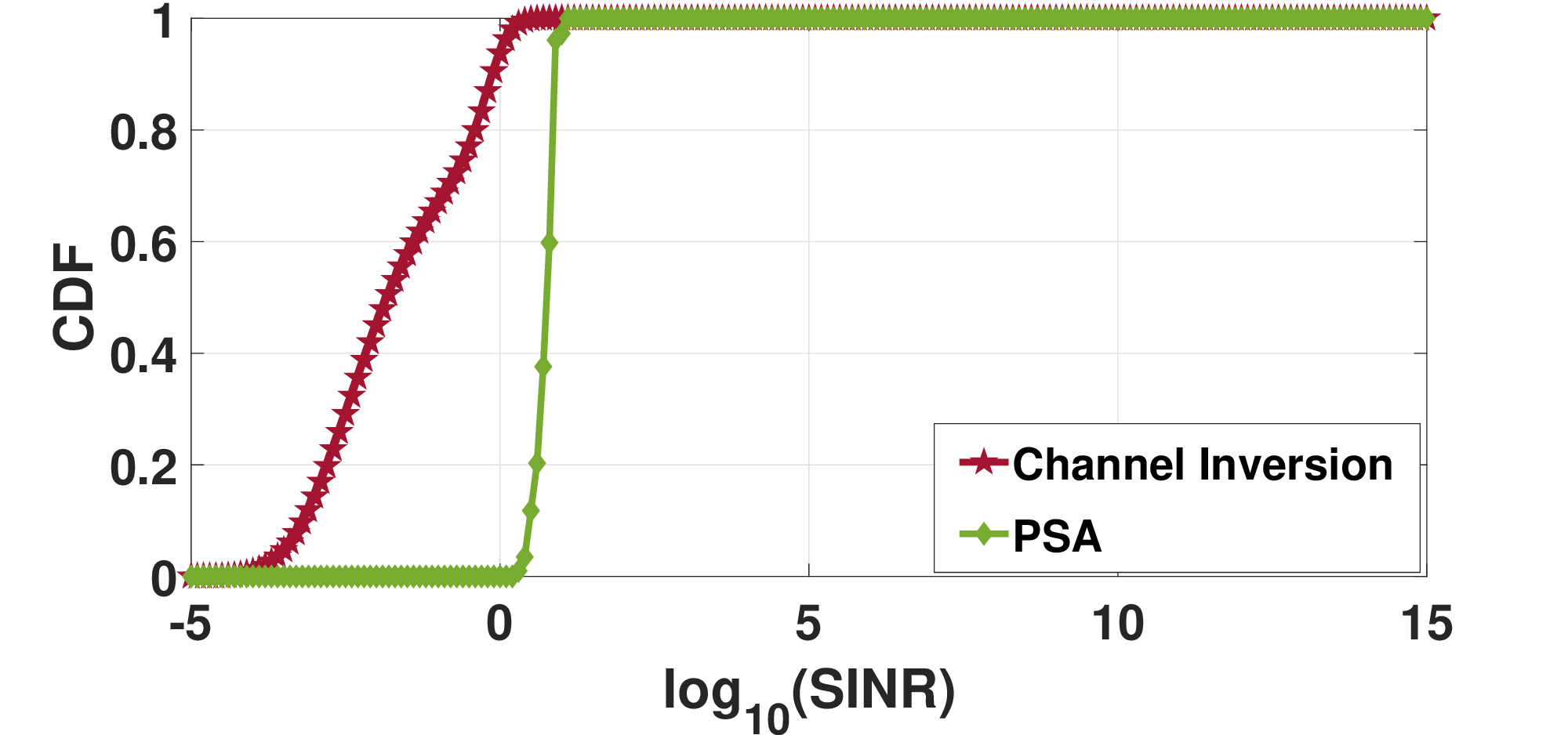}
	\caption{CDF of the minimum achievable downlink SINRs for different power control strategies in $64$ UEs case.}
	\label{fig:e9}
\end{figure}

We now compare the performance of the PSA-based power control algorithm against channel inversion-based power control for a $256$ AP and $64$ UE system. In Fig.~\ref{fig:e8} and Fig.~\ref{fig:e9}, we plot the CDFs of the minimum achievable rates for uplink and downlink, respectively. We observe that PSA-based power control provides about 60\% improvement for the 90\% likely minimum achievable rate of the considered system in uplink and more than doubles the same for in the downlink, as compared to channel inversion-based power control. 


\begin{figure}[t]
	\centering
	\includegraphics[width=27em]{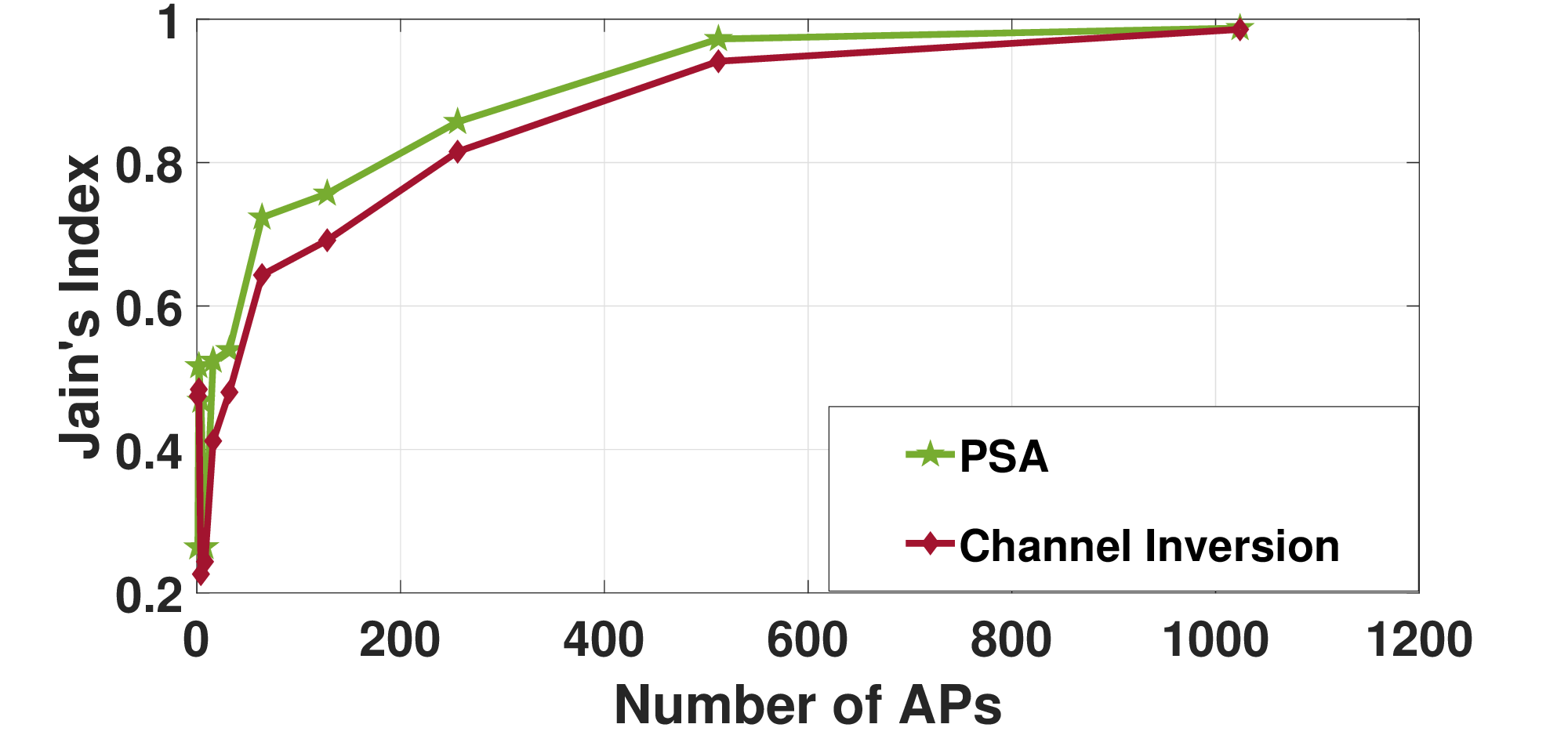}
	\caption{Comparison of Jain's Fairness Index  in the uplink for $64$ UEs case.}
	\label{fig:e10}
\end{figure}

\begin{figure}[t]
	\centering
	\includegraphics[width=27em]{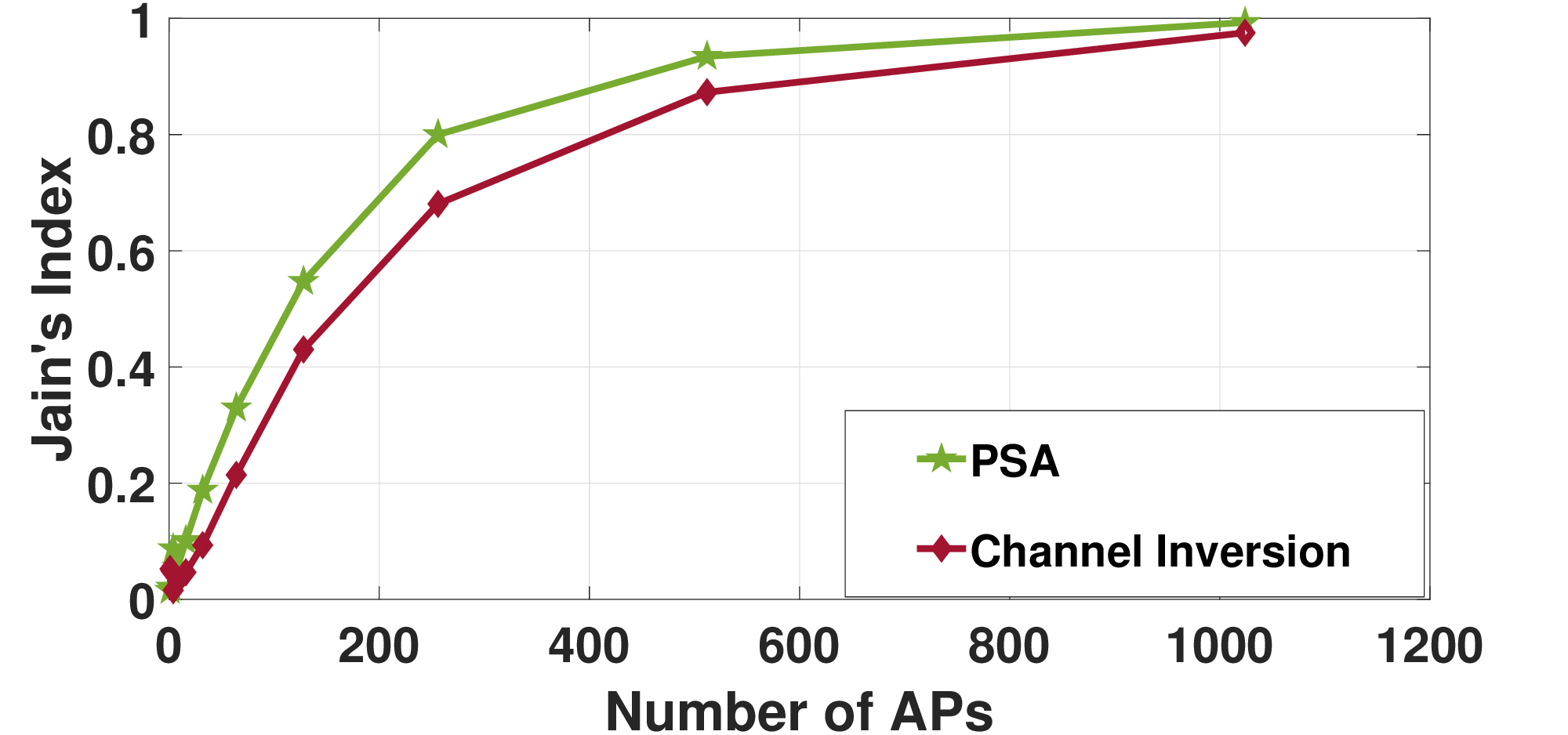} 
	\caption{Comparison of Jain's Fairness Index  in downlink for $64$ UEs case. }
	\label{fig:e11}
\end{figure}

Finally, to quantify the improvement in fairness, we compare Jain's Fairness index between the PSA and channel inversion-based power control as a function of the number of APs for uplink and downlink in Fig.~\ref{fig:e10} and Fig.~\ref{fig:e11} respectively. We observe that the use of PSA-based power control results in as much as 50\% improvement in the fairness of the system for lower AP densities.


\section{Conclusion}
In this paper, we discussed the performance of the CF mMIMO system under a probabilistic LoS/NLoS channel using different power control strategies at moderate AP densities. First, we evaluated the performance of our proposed system with Channel inversion-based power control and then compared its performance with max-min fairness-based power control. Motivated by the performance gap, we formulated the PSA-based power control to provide approximately uniform coverage to all the UEs. Future work could include the design and use of other meta-heuristic algorithm-based power control strategies. \\

\large{\textbf{References:}} \\
\small
$\left[1\right]$  Marzetta, Thomas L.: `Fundamentals of massive MIMO', \textit{Cambridge University Press}, 2016 \\
$\left[2\right]$  Chopra, Ribhu and Murthy, Chandra R and Papazafeiropoulos, Anastasios K.: `Uplink performance analysis of cell-free mMIMO systems under channel aging', \textit{IEEE Communications Letters}, 2021, \textbf{25}, pp.~2206--2210 \\
$\left[3\right]$ Bj{\"o}rnson, Emil and Sanguinetti, Luca and Wymeersch, Henk and Hoydis, Jakob and Marzetta, Thomas L.: `Massive MIMO is a reality—What is next?: Five promising research directions for antenna arrays', \textit{Elsevier, Digital Signal Processing}, 2019, \textbf{94}, pp.~3--20 \\
$\left[4\right]$ Panwar, Nisha and Sharma, Shantanu and Singh, Awadhesh Kumar.: `A survey on 5G: The next generation of mobile communication', \textit{Elsevier ,Physical Communication}, 2016, \textbf{18}, pp.~64--84 \\
$\left[5\right]$ Ngo, Hien Quoc and Tran, Le-Nam and Duong, Trung Q and Matthaiou, Michail and Larsson, Erik G.: `On the total energy efficiency of cell-free massive MIMO', \textit{IEEE Transactions on Green Communications and Networking}, 2017, \textbf{2}, pp.~25--39 \\
$\left[6\right]$ Yang, Ping and Xiao, Yue and Xiao, Ming and Li, Shaoqian.: `6G wireless communications: Vision and potential techniques', \textit{IEEE network}, 2019, \textbf{33}, pp.~70--75 \\
$\left[7\right]$ Larsson, Erik G and Edfors, Ove and Tufvesson, Fredrik and Marzetta, Thomas L.: `Massive MIMO for next generation wireless systems', \textit{IEEE communications magazine}, 2014, \textbf{52}, pp.~186--195 \\
$\left[8\right]$ Nayebi, Elina and Ashikhmin, Alexei and Marzetta, Thomas L and Yang, Hong.: `Cell-free massive MIMO systems', \textit{Asilomar Conference on Signals, Systems and Computers}, 2015, \textbf{49}, pp.~695--699 \\
$\left[9\right]$ 	Ngo, Hien Quoc and Ashikhmin, Alexei and Yang, Hong and Larsson, Erik G and Marzetta, Thomas L.: `Cell-free massive MIMO versus small cells', \textit{IEEE Transactions on Wireless Communications}, 2017, \textbf{16}, pp.~1834--1850 \\
$\left[10\right]$ Ngo, Hien Quoc and Ashikhmin, Alexei and Yang, Hong and Larsson, Erik G and Marzetta, Thomas L.: `Cell-free massive MIMO: Uniformly great service for everyone', \textit{IEEE international workshop on signal processing advances in wireless communications (SPAWC)}, 2015, \textbf{16}, pp.~201--205 \\
$\left[11\right]$ 	{{\"O}zdogan, {\"O}zgecan and Bj{\"o}rnson, Emil and Zhang, Jiayi}.: `Performance of cell-free massive MIMO with Rician fading and phase shifts', \textit{IEEE Transactions on Wireless Communications}, 2019, \textbf{18}, pp.~5299--5315 \\
$\left[12\right]$ Zhang, Yao and Zhou, Meng and Cao, Haotong and Yang, Longxiang and Zhu, Hongbo.: `On the performance of cell-free massive MIMO with mixed-ADC under Rician fading channels', \textit{IEEE Communications Letters}, 2019, \textbf{24}, pp.~43--47 \\
$\left[13\right]$ 	Jin, Si-Nian and Yue, Dian-Wu and Nguyen, Ha H.: `Spectral and energy efficiency in cell-free massive MIMO systems over correlated Rician fading', \textit{IEEE Systems Journal}, 2020, \textbf{15}, pp.~2822--2833 \\
$\left[14\right]$ Nayebi, Elina and Ashikhmin, Alexei and Marzetta, Thomas L and Yang, Hong and Rao, Bhaskar D.: `Precoding and power optimization in cell-free massive MIMO systems', \textit{IEEE Transactions on Wireless Communications}, 2017, \textbf{16}, pp.~4445--4459 \\
$\left[15\right]$ Bashar, Manijeh and Cumanan, Kanapathippillai and Burr, Alister G and Ngo, Hien Quoc and Debbah, M{\'e}rouane.: `Cell-free massive MIMO with limited backhaul', \textit{IEEE International Conference on Communications (ICC)}, 2018, \textbf{64}, pp.~1--7 \\
$\left[16\right]$ Papazafeiropoulos, Anastasios and Kourtessis, Pandelis and Di Renzo, Marco and Chatzinotas, Symeon and Senior, John M.: `Performance analysis of cell-free massive MIMO systems: A stochastic geometry approach', \textit{IEEE Transactions on Vehicular Technology}, 2020, \textbf{69}, pp.~3523--3537 \\
$\left[17\right]$ Zhang, Jiayi and Wei, Yinghua and Bj{\"o}rnson, Emil and Han, Yu and Jin, Shi.: `Performance analysis and power control of cell-free massive MIMO systems with hardware impairments', \textit{IEEE Access}, 2018, \textbf{6}, pp.~55302--55314 \\
$\left[18\right]$ 	{Bj{\"o}rnson, Emil and Sanguinetti, Luca}.: `Making cell-free massive MIMO competitive with MMSE processing and centralized implementation', \textit{IEEE Transactions on Wireless Communications}, 2019, \textbf{19}, pp.~77--90 \\
$\left[19\right]$ 	Atzeni, Italo and Arnau, Jesús and Kountouris, Marios.: `Downlink Cellular Network Analysis With LOS/NLOS Propagation and Elevated Base Stations', \textit{IEEE Transactions on Wireless Communications}, 2017, \textbf{17}, pp.~142--156 \\
$\left[20\right]$ 	Mukherjee, Sudarshan and Chopra, Ribhu.: `Performance analysis of cell-free massive MIMO systems in LoS/NLoS channels', \textit{IEEE Transactions on Vehicular Technology}, 2022, \textbf{71}, pp.~6410--6423 \\
$\left[21\right]$ 	{{\"O}zdogan, {\"O}zgecan and Bj{\"o}ornson, Emil and Zhang, Jiayi}.: `Cell-free massive MIMO with Rician fading: Estimation schemes and spectral efficiency', \textit{Asilomar Conference on Signals, Systems, and Computers}, 2018, \textbf{52}, pp.~975--979 \\
$\left[22\right]$ 	Farooq, Muhammad and Ngo, Hien Quoc and others.: `A low-complexity approach for max-min fairness in uplink cell-free massive MIMO', \textit{IEEE Vehicular Technology Conference (VTC2021-Spring)}, 2021, \textbf{93}, pp.~1--6 \\
$\left[23\right]$ Tran, Le-Nam and Ngo, Hien Quoc.: `First-order methods for energy-efficient power control in cell-free massive MIMO', \textit{Asilomar Conference on Signals, Systems, and Computers}, 2019, \textbf{53}, pp.~848--852 \\
$\left[24\right]$ Bashar, Manijeh and Akbari, Ali and Cumanan, Kanapathippillai and Ngo, Hien Quoc and Burr, Alister G and Xiao, Pei and Debbah, Merouane and Kittler, Josef.: `Exploiting deep learning in limited-fronthaul cell-free massive MIMO uplink', \textit{IEEE Journal on Selected Areas in Communications}, 2020, \textbf{38}, pp.~1678--1679 \\
$\left[25\right]$ Saray, Akbar Mazhari and Ebrahimi, Afshin.: `MAX-MIN Power Control of Cell-Free Massive MIMO System employing Deep Learning', \textit{West Asian Symposium on Optical and Millimeter-wave Wireless Communications (WASOWC)}, 2022, \textbf{4}, pp.~1--4 \\
$\left[26\right]$ Rajapaksha, Nuwanthika and Manosha, KB Shashika and Rajatheva, Nandana and Latva-Aho, Matti.: `Deep learning-based power control for cell-free massive MIMO networks', \textit{ICC IEEE International Conference on Communications}, 2021, \textbf{21}, pp.~1--7 \\
$\left[27\right]$ Chakraborty, Sucharita and Bj{\"o}rnson, Emil and Sanguinetti, Luca.: `Centralized and distributed power allocation for max-min fairness in cell-free massive MIMO', \textit{asilomar conference on signals, systems, and computers}, 2019, \textbf{53}, pp.~576--580 \\
$\left[28\right]$ Conceição, Filipe and Antunes, Carlos Henggeler and Gomes, Marco and Silva, Vitor and Dinis, Rui.: `Max-Min Fairness Optimization in Uplink Cell-Free Massive MIMO Using Meta-Heuristics', \textit{IEEE Transactions on Communications}, 2022, \textbf{70}, pp.~1792--1807 \\
$\left[29\right]$ Anubhab Chowdhury and Pradip Sasmal and	Chandra R. Murthy and Ribhu Chopra.: `On the Performance of Distributed Antenna Array Systems With Quasi-Orthogonal Pilots', \textit{{IEEE} Trans. Veh. Technol.}, 2022, \textbf{71}, pp.~3326--3331 \\
$\left[30\right]$ Kim, Dongsun and Lee, Jemin and Quek, Tony QS.: `Multi-layer unmanned aerial vehicle networks: Modeling and performance analysis', \textit{IEEE Transactions on Wireless Communications}, 2019, \textbf{19}, pp.~325--339 \\
$\left[31\right]$ Mondal, Bishwarup and Thomas, Timothy A. and Visotsky, Eugene and Vook, Frederick W. and Ghosh, Amitava and Nam, Young-han and Li, Yang and Zhang, Jianzhong and Zhang, Min and Luo, Qinglin and Kakishima, Yuichi and Kitao, Koshiro.: `3D channel model in 3GPP', \textit{IEEE Communications Magazine}, 2015, \textbf{53}, pp.~16--23 \\
$\left[32\right]$ 	Kennedy, James and Eberhart, Russell.: `Particle swarm optimization', \textit{Proceedings of ICNN international conference on neural networks}, 2015, \textbf{4}, pp.~1942--1948 \\
$\left[33\right]$ 	Blackwell, Tim.: `Particle swarm optimization in dynamic environments', \textit{Evolutionary computation in dynamic and uncertain environments}, 2007, \textbf{4}, pp.~29--49\\
$\left[34\right]$ Parida, Priyabrata and Dhillon, Harpreet S and Molisch, Andreas F.: `Downlink performance analysis of cell-free massive MIMO with finite fronthaul capacity', \textit{IEEE 88th Vehicular Technology Conference (VTC-Fall)}, 2018, pp.~1--6\\
$\left[35\right]$ 	Zhang, Yao and Cao, Haotong and Zhou, Meng and Yang, Longxiang.: `Cell-free massive MIMO: Zero forcing and conjugate beamforming receivers', \textit{Journal of Communications and Networks}, 2007, \textbf{21}, pp.~529--538\\


\end{document}